\documentclass[twocolumn]{aastex61}

\received{}
\revised{}
\accepted{}
\submitjournal{ApJ}

\shorttitle{The Structures of Very Massive Galaxies at $1.5<z<3$}
\shortauthors{Marsan et al.}

\begin{document}

\title{HST F160W Imaging of Very Massive Galaxies at $1.5<z<3.0$: Diversity of Structures and the Effect of Close Pairs on Number Density Estimates}

\correspondingauthor{Z. Cemile Marsan}
\email{cmarsan@yorku.ca}

\author{Z. Cemile Marsan}
\altaffiliation{York Science Fellow}
\affil{Department of Physics and Astronomy, York University, 
4700 Keele Street, Toronto, Ontario, MJ3 1P3, Canada}

\author{Danilo Marchesini}
\affil{Department of Physics and Astronomy, Tufts University, 
574 Boston Avenue, Medford, MA 02155, USA}

\author{Adam Muzzin}
\affil{Department of Physics and Astronomy, York University, 
4700 Keele Street, Toronto, Ontario, MJ3 1P3, Canada}

\author{Gabriel B. Brammer}
\affiliation{Space Telescope Science Institute, 3700 San Martin Drive, Baltimore, MD 21218, USA}
\affiliation{Cosmic Dawn Center, Niels Bohr Institute, University of Copenhagen, Juliane Maries Vej 30, DK-2100 Copenhagen \O, Denmark}

\author{Rachel Bezanson}
\affiliation{Department of Physics and Astronomy and PITT PACC, 
University of Pittsburgh, Pittsburgh, PA 15260, USA}

\author{Marijn Franx}
\affiliation{Leiden Observatory, Leiden University, 
PO Box 9513, 2300 RA Leiden, The Netherlands}

\author{Ivo Labb\'e}
\affiliation{Centre for Astrophysics and Supercomputing, Swinburne University of Technology, 
Hawthorn, VIC 3122, Australia}

\author{Britt Lundgren}
\affiliation{Department of Physics, University of North Carolina Asheville, 
One University Heights, Asheville, NC 28804, USA}

\author{Gregory Rudnick}
\affil{Department of Physics and Astronomy, The University of Kansas, 
1251 Wescoe Hall Drive, Lawrence, KS 66045, USA}

\author{Mauro Stefanon}
\affiliation{Leiden Observatory, Leiden University, 
PO Box 9513, 2300 RA Leiden, The Netherlands}

\author{Pieter van Dokkum}
\affil{Department of Astronomy, Yale University, New Haven, CT 06511, USA}

\author{David Wake}
\affiliation{Department of Physics, University of North Carolina Asheville, 
One University Heights, Asheville, NC 28804, USA.}
\affiliation{Department of Physical Sciences, The Open University, Milton Keynes MK7 6AA, UK}

\author{Katherine E. Whitaker}
\affiliation{Department of Physics, University of Connecticut, Storrs, CT 06269, USA}
\affiliation{Cosmic Dawn Center (DAWN), Niels Bohr Institute, University of Copenhagen / DTU-Space, Technical University of Denmark}








\begin{abstract}
We present a targeted follow-up \textit{Hubble Space Telescope} WFC3  F160W imaging study of very massive galaxies 
($\log(M_{\rm{star}}/M_{\odot})> 11.2$) selected from a combination of ground-based 
near-infrared galaxy surveys (UltraVISTA, NMBS-II, UKIDSS UDS) at $1.5<z<3$. 
We find that these galaxies are diverse in their structures, with $\sim1/3$ of the targets being composed of close pairs, and span a wide range in sizes. 
At $1.5<z<2.5$, the sizes of both star-forming and quiescent galaxies 
are consistent with the extrapolation of
the stellar mass-size relations determined at lower stellar masses. At $2.5<z<3.0$, however, we find evidence that quiescent galaxies are systematically 
larger than expected based on the extrapolation of the relation derived using lower stellar mass galaxies.
We used the observed light profiles of the blended systems to decompose their stellar masses and investigate 
the effect of the close pairs on the measured number densities of very massive galaxies in the early universe. 
We estimate correction factors to account for close-pair blends and apply them to the observed stellar mass functions measured using ground-based surveys. 
Given the large uncertainties associated with this extreme population of galaxies, there is currently little tension between the (blending-corrected) number density estimates and predictions from theoretical models. Although we currently lack the statistics to robustly correct for close-pair blends, we show that this is a systematic effect which can reduce the observed number density of very massive galaxies by up to a factor of $\sim1.5$, and should be accounted for in future studies of stellar mass functions. 

\end{abstract}

\keywords{galaxies}



\section{Introduction} \label{sec3-intro}

In contrast to the hierarchical assembly of dark matter haloes, observations indicate that the most massive galaxies in the nearby universe were among the first to build-up their stellar mass and quench.  In the nearby universe, massive galaxies are found to be older, more metal rich and to have formed their stars more rapidly and at earlier cosmic epochs compared to their lower-mass counterparts   
\citep{terlevich01, bernardi03, trager00, thomas05, gallazzi05, gallazzi06, yamada06, kuntschner10, mcdermid15}.
Corroborating their early formation times are results from recent deep near-infrared (NIR) surveys which reveal that very massive galaxies were already in place by $z\sim4$ (merely $\sim1.5$~Gyr after the Big Bang; e.g., 
\citealt{marchesini10, ilbert13, muzzin13a, straatman14, duncan14, tomczak14, caputi15, grazian15, song16, davidzon17}), 
and spectroscopic follow-up campaigns, confirming that these massive galaxies have evolved stellar populations at $z>3$ 
(e.g., \citealt{marsan15, marsan17, glazebrook17, schreiber18}). Thus, the observed properties of the most massive galaxies serve as critical benchmarks to understand the detailed physical mechanisms that impact galaxy formation and evolution in the early universe. 

A two-phase scenario has been proposed for the evolution of massive galaxies: a rapid, compact formation at early epochs via highly dissipative processes (e.g. by experiencing gas-rich major mergers or violent disk instabilities; \citealt{hopkins06, dekel09, krumholz10, dekel14, wellons15, bournaud16}), and following the quenching of star-formation, a later phase of assembly dominated by undergoing dry minor mergers with satellite galaxies \citep{nipoti03, khochfar06,naab09b, oser10, hilz12, hilz13}. 
Several observables serve to corroborate this scenario: the uniform, old stellar populations of $z\sim0$ massive galaxies \citep{mcdermid15}, the build-up of stellar haloes in (central) massive galaxies (e.g. \citealt{buitrago17, huang18a, huang18b}), 
and the dramatic size evolution observed for the massive, quiescent galaxy population since $z\sim2$ 
\citep{trujillo06, buitrago08, franx08, vandokkum08b, cimatti08, bezanson09, damjanov09, kriek09b, williams10, vandokkum10a, vanderwel11, newman12, szomoru12, whitaker12a, patel13, vanderwel14a, belli14a, belli15, hill17}.

The structural evolution of galaxies is sensitive to their assembly history and feedback processes, as such, the observed size and morphology of galaxies in various environment and halo mass regimes is a critical benchmark for theoretical models to reproduce (e.g., \citealt{genel17, furlong17}). A census of galaxy size has now been obtained out to $z\sim4$ across a wide range in stellar mass and star formation activity 
(e.g., \citealt{shen03, trujillo04, bezanson09, patel13, vandokkum14, vanderwel14a, straatman15, allen17}). 
However, the majority of information on the size evolution of \textit{massive} galaxies is obtained from samples with stellar masses in the range of $1-2 \times 10^{11} M_{\odot}$; as such, the size-mass relation at the extreme massive end of the galaxy population (i.e., $\log(M_{*}/M_{\odot}$)$\geq11.25$) at $z>1.5$ remains poorly constrained.
Abundance matching techniques suggest that ultra-massive galaxies, those with $\log(M_{*}/M_{\odot})>11.60$ should reside in dark matter haloes of a few $\times 10^{14} M_{\odot}$ at all redshifts, implying that they are the progenitors of the Brightest Cluster Galaxies (BCGs) in the local universe. Therefore, measuring how these massive systems evolve in size compared to their (relatively) lower-mass cousins could provide valuable information on how their assembly takes place, and whether this evolution is related to their halo properties (e.g., concentration, mass, or subhalo occupation number).


Owing to the low spatial density of these objects, 
identifying a statistically large sample of very massive galaxies requires relatively deep and wide NIR surveys using ground-based facilities, which typically lack the spatial resolution to derive robust sizes for these compact, distant galaxies (the typical FWHM $\sim~0.8-1^{''}$ corresponds to a physical size of $\sim~6-9$~kpc at $z=1.5-3$). To this end, we have obtained follow-up \textit{HST}/WFC3 $H_{160}$ imaging for a sample of very massive ($\log(M_{*}/M_{\odot})>11.25$) galaxies at $1.5<z<3.0$ selected using relatively deep and wide-field ground based NIR surveys. The $H_{160}$ band, the reddest filter currently available for high-resolution imaging, probes the rest-frame wavelength regime just blueward of the $r$ band ($\sim 6100$ {\AA}) at $z\sim1.5$ to wavelengths just redward of the rest-frame Balmer break at $z\sim3.0$ (i.e., $\sim~3900$~{\AA})

In this study, we present the \textit{HST}/WFC3 $H_{160}$ imaging for 37 targets with stellar masses $\log(M_{*}/M_{\odot})>11.2$ at $1.5<z<3.0$ in the NMBS-II, UltraVISTA and UKIDSS UDS. In Section~\ref{sec3-data} we briefly describe the datasets used to select this sample and the targeted \textit{HST} observations. Section~\ref{sec3-analysis} presents the analysis and relevant measurements employed in this study. 
We present the results in Section~\ref{sec3-results} and summarize these results in Section~\ref{sec3-summary}. 
Throughout this paper we assume the standard $\Lambda$CDM cosmological parameters $\Omega{_M}=0.3$, $\Omega{_\Lambda}=0.7$ with $H_{0}=70$~km~s$^{-1}$~Mpc$^{-1}$ and a \citet{chabrier03} initial mass function (IMF). All magnitudes listed are in the AB system.

\section{Data} \label{sec3-data}

\subsection{Parent Catalogs} \label{sec3-parencat}

We use the NEWFIRM Medium Band Survey-II (NMBS-II, Annunziatella et al., in prep) and the UltraVISTA survey \citep{muzzin13a} to identify and select the rare, very massive ($\log(M_{*}/M_{\odot})> 11.4$) galaxies at $1.5<z<3.0$ for targeted follow-up \textit{HST}~WFC3/$H_{160}$ imaging (GO12990, PI: Muzzin). 

We also utilize the \textit{HST}~$H_{160}$ imaging follow-up study of $\log(M_*/M_{\odot})> 11.25$ quiescent galaxies at $2.5<z<3.0$ (GO13002, PI: Williams) selected from the UKIRT Infrared Deep Sky Survey (UKIDSS) Ultra-Deep Survey (UDS; \citealt{lawrence07}) to extend our sample to include massive, quiescent galaxies. These surveys combine to an effective area of $\sim5.9$~deg$^2$. Below, we briefly describe the photometric catalogs and the spectral energy distribution (SED) fitting, and refer the reader to the works mentioned for further details related to data processing, photometry and SED modeling assumptions.   

The NMBS-II is a wide, but relatively shallow NIR ($K=21.75$, 5$\sigma$) survey, covering a total area of $\sim4.25$~deg$^2$ in the CFHTLS-D1, CFHTLS-D4, COSMOS and MUSYC fields. 
This survey combines deep NIR medium-bandwidth photometry ($J_{1}$, $J_{2}$, $J_{3}$, $H_{1}$, $H_{2}$) with the existing UV, optical and NIR data in these fields to accurately identify evolved, massive galaxies by tracing the rest-frame optical break ($\sim4000${\AA}) at $z>1$. 
In the COSMOS field \citep{scoville07}, where the NMBS-II footprint overlaps with the UltraVISTA survey ($\sim1.62$~deg$^2$, $K=23.8$, \citealt{mccracken12}) we used the $K_{S}$-selected  galaxy catalog from \citet{muzzin13a} to complement the wider-field, yet shallower, NMBS-II dataset. Photometric redshifts are estimated using EAZY \citep{brammer08} and the stellar population parameters, including stellar mass, are calculated using FAST \citep{kriek09a} assuming exponentially declining star formation histories, fixed solar metallicity and the \citet{calzetti00} dust reddening law.
 
Targets in the UDS field are selected from the photometric catalog presented in \citet{williams09, williams10} and \citet{quadri12} using Data Release 8 of the UKIDSS NIR imaging (reaching 5$\sigma$ point-source depth of $K=24.5$) over an effective area of $\sim 0.62$~deg$^{2}$. This dataset also includes $u^{'},B, V,R, i^{'}, z^{'}, J,H,$ and Spitzer IRAC $3.6~\mu$m and $4.5~\mu$m band photometry. Photometric redshifts, stellar masses and other stellar population parameters were estimated in an identical manner as described above for the NMBS-II and UltraVISTA photometric catalog papers. 

\subsection{Targeted sample for HST WFC3 H$_{160}$ Imaging} \label{sec3-hst}

\begin{figure}[!h]
\epsscale{1.1}
\plotone{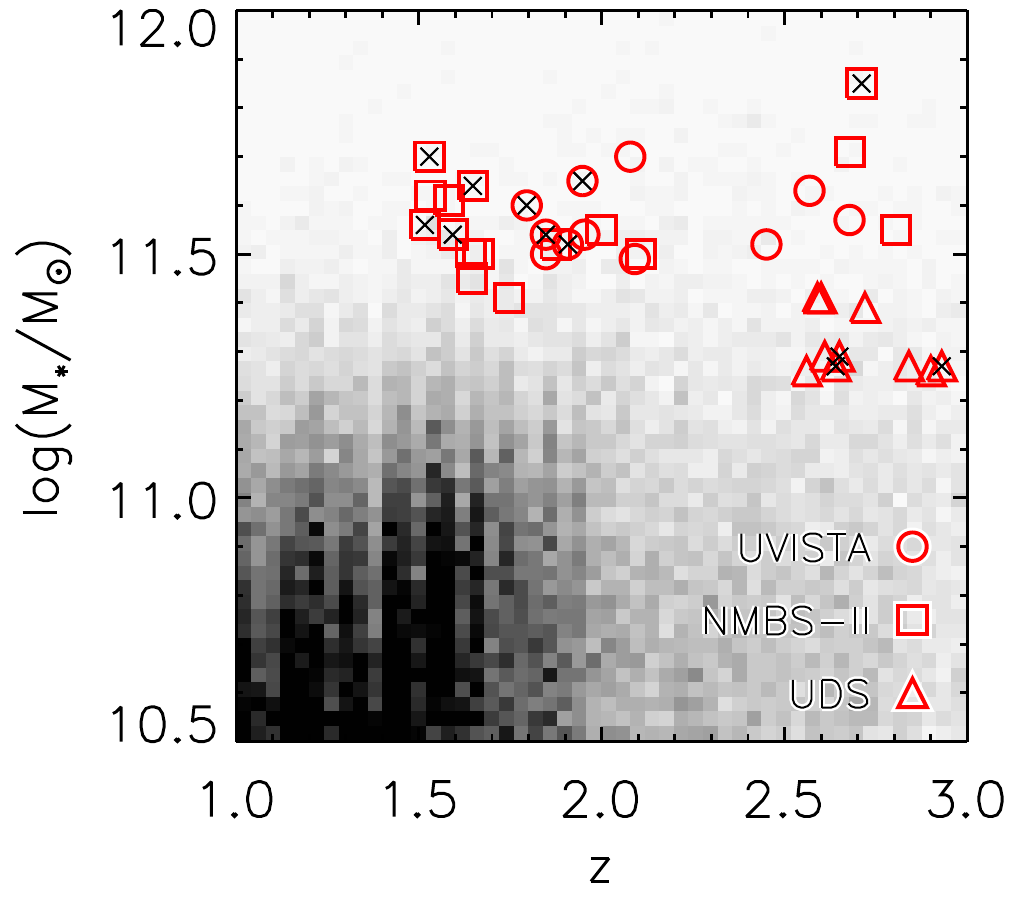}
\caption{The distribution of stellar masses as a function of redshift for the sample of very massive galaxies at $1.5<z<3.0$ targeted with HST WFC3 $H_{160}$ band imaging (open red symbols). Targets selected from the UltraVISTA D1, NMBS-II and UDS DR8 catalogs are represented as circles, squares and triangles, respectively. Targets that are discovered to be close-pairs are marked with X. The grayscale represents the distribution of galaxies above the magnitude completeness limit of each parent photometric galaxy survey.} \label{fig3-parentcat}
\end{figure}

The point spread function (PSF) of typical ground-based near-IR imaging is insufficient to reliably measure sizes of the smallest galaxies at $z >1$ (where FWHM $\sim0.^{''}8$ corresponds to physical distances of $\sim 6-7$~kpc at $1<z<3$ ). We therefore utilized follow-up \textit{HST} imaging in the reddest WFC3 band, $H_{160}$, to obtain size measurements of our targets.

Figure~\ref{fig3-parentcat} highlights that the galaxies in this sample are among the most massive ones at the epochs probed. The grayscale representation shows all galaxies brighter than the magnitude limit of each parent catalog (UDS DR8: $K=24$; UltraVISTA DR1: $K=23.8$; NMBS-II: $K=21.75$), with red symbols denoting the follow-up targets. 
From the combined UltraVISTA DR1 and NMBS-II photometric catalogs, a total of 27 targets at $1.5 < z < 3$ with robust stellar mass estimates $\log(M{_*}/M_{\odot})> 11.4$ were selected for \textit{HST}/WFC3 follow-up observations (GO12990, PI: Muzzin); represented with open circles and squares respectively in Figure~\ref{fig3-parentcat}. 
The open triangles represent the additional ten quiescent galaxies with $\log(M{_*}/M_{\odot})> 11.25$ at $2.5 < z < 3.0$  that were selected for \textit{HST}/WFC3 follow-up observations in the UDS field (GO13002, PI: Williams; see \citealt{patel17}). The combination of these data sets yields a total of 37 galaxies at  $1.5<z<3$  with  $\log(M{_*}/M_{\odot})> 11.25$, increasing the available high-resolution imaging for this extreme population by a factor of $\sim2$ compared to the CANDELS dataset for galaxies with $\log(M{_*}/M_{\odot})> 11.4$ \citep{vanderwel14a}.

\begin{figure*}
\epsscale{0.75}
\plotone{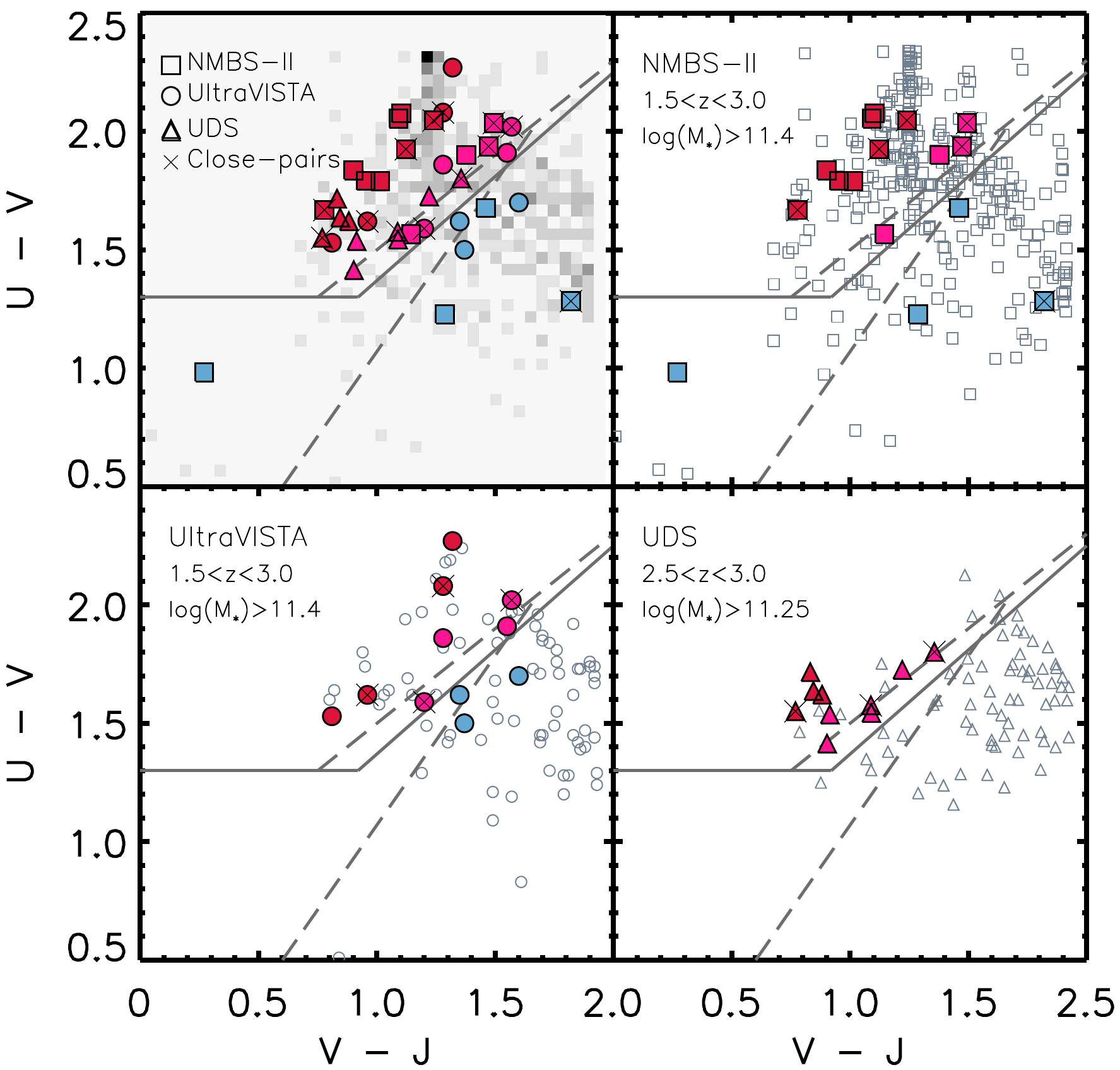}
\caption{Rest-frame $U-V$ versus $V-J$ color-color diagram. The filled symbols indicate the rest-frame colors for the $1.5<z<3.0$ galaxies with targeted \textit{HST} observations. Targets that are discovered to be close-pairs are marked with X. The cuts used to separate star-forming (blue) from quiescent (red and pink) galaxies from \citet{vanderwel14a} are shown with solid gray lines. Also shown, with dashed lines, is the separation between quiescent, unobscured and dusty star-forming galaxies from \citet{martis16}. \textit{Top left panel} displays the colors for the targeted \textit{HST} sample, along with all sources that satisfy the sample selection criteria in each photometric galaxy catalog (grayscale). \textit{Top right and bottom panels} focus on the individual parent photometric galaxy catalogs and display the colors for all sources that satisfy the redshift and stellar mass criteria adopted (indicated in legend).} \label{fig3-uvj}
\end{figure*}

Figure~\ref{fig3-uvj} shows the rest-frame $U-V$ vs. $V-J$ color diagram, frequently used to distinguish star-forming and quiescent galaxies  (e.g., \citealt{labbe05}, \citealt{whitaker11}, \citealt{muzzin13b}). 
The rest-frame colors were calculated using EAZY \citep{brammer08}. For consistency with \citet{vanderwel14a}, we used the updated rest-frame color cuts of \citet{williams09} to separate star-forming (blue symbols) from quiescent galaxies (red and pink symbols). Based on their rest-frame colors, 30 ($\sim80\%$ of total) very massive galaxies at $1.5<z<3$ fall into the quiescent region. Compared to the quiescent fractions derived by \citet{martis16} using combination of the UltraVISTA DR1 and CANDELS datasets, our sample is characterized by a larger quiescent fraction, although the estimated quiescent fraction in the largest stellar mass bin ($11.5<\log(M_{*}/M_{\odot})<11.8$) in \citet{martis16} are very uncertain.
Noticing that a significant portion of our targets lie close to the \textit{UVJ} selection cuts, 
we calculated a quiescent fraction to account for contamination from potential post-starburst or fading galaxies with intermediate colors (pink symbols in Figure~\ref{fig3-uvj}). Selecting galaxies that are $>0.2$~mag away from the 
diagonal color cut (red symbols in Figure~\ref{fig3-uvj}), we calculated a conservative quiescent fraction of $\sim45\%$. 
The few star-forming galaxies tend to have colors consistent with accumulating along the 
quiescent-star forming transition zone at the dusty end of the star-forming region, with only one relatively unobscured star-forming galaxy. 

\begin{figure*}[!h]
\epsscale{1.0}
\plotone{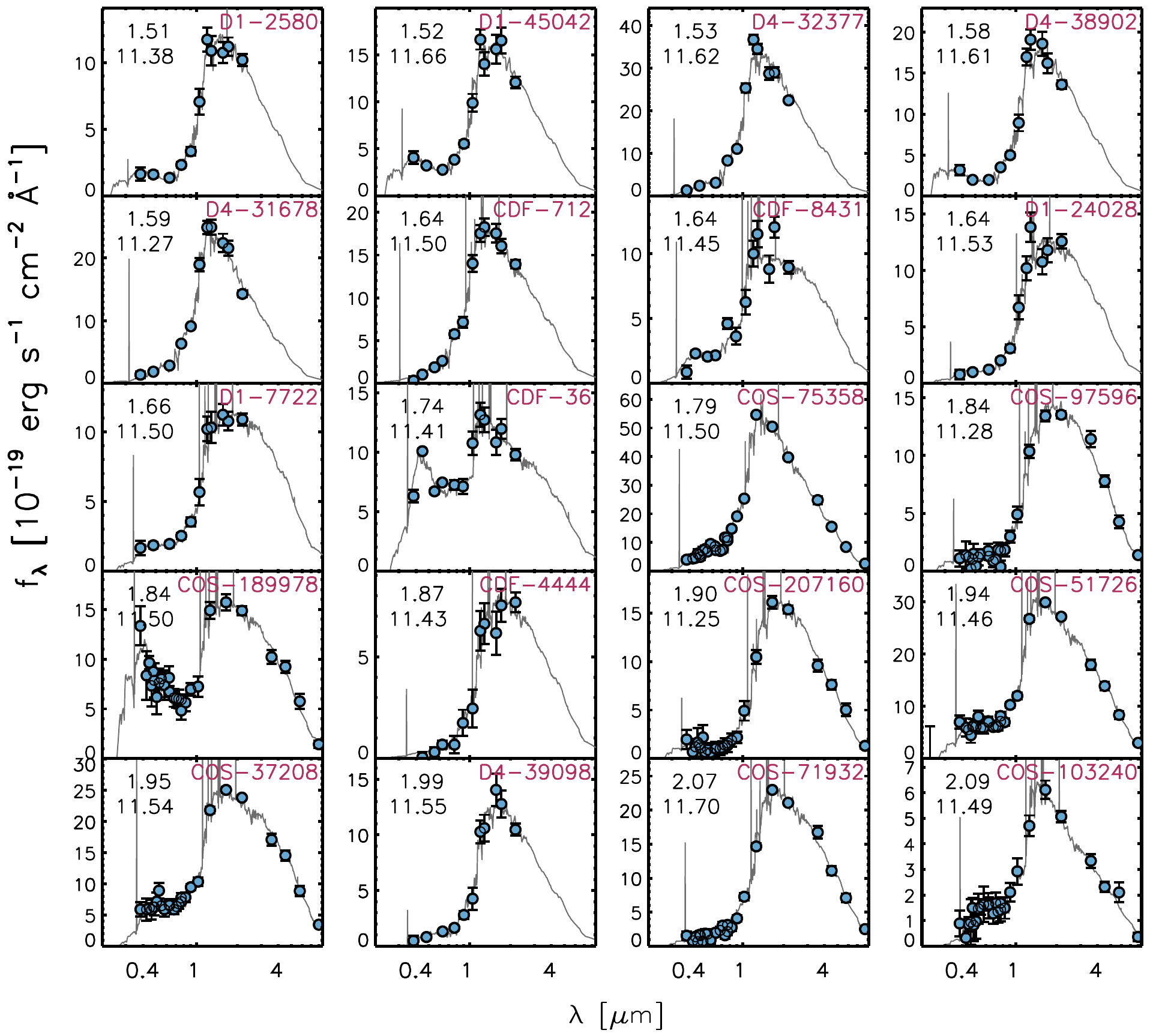}
\caption{The observed UV-IR spectral energy distributions for 20 of the 37 massive galaxies at $1.5<z<3$ selected for targeted follow-up observations. 
Photometry from parent catalogs are shown in blue filled symbols, in units of $10^{-19}$ erg~cm$^{-2}$~s$^{-1}$~\AA$^{-1}$
and the gray curves represent the best-fit EAZY templates. The ID, $z_{phot}$ and stellar mass ($\log(M_{*}/M_{\odot})$) of targets are listed in each panel, using the abbreviations COS, D1, D4 and CDF to denote targets in the COSMOS, CFHTD-1, CFHTD-4 and ECDFS fields, respectively.  } \label{fig-sed1}
\end{figure*}

\begin{figure*}
\epsscale{1.0}
\plotone{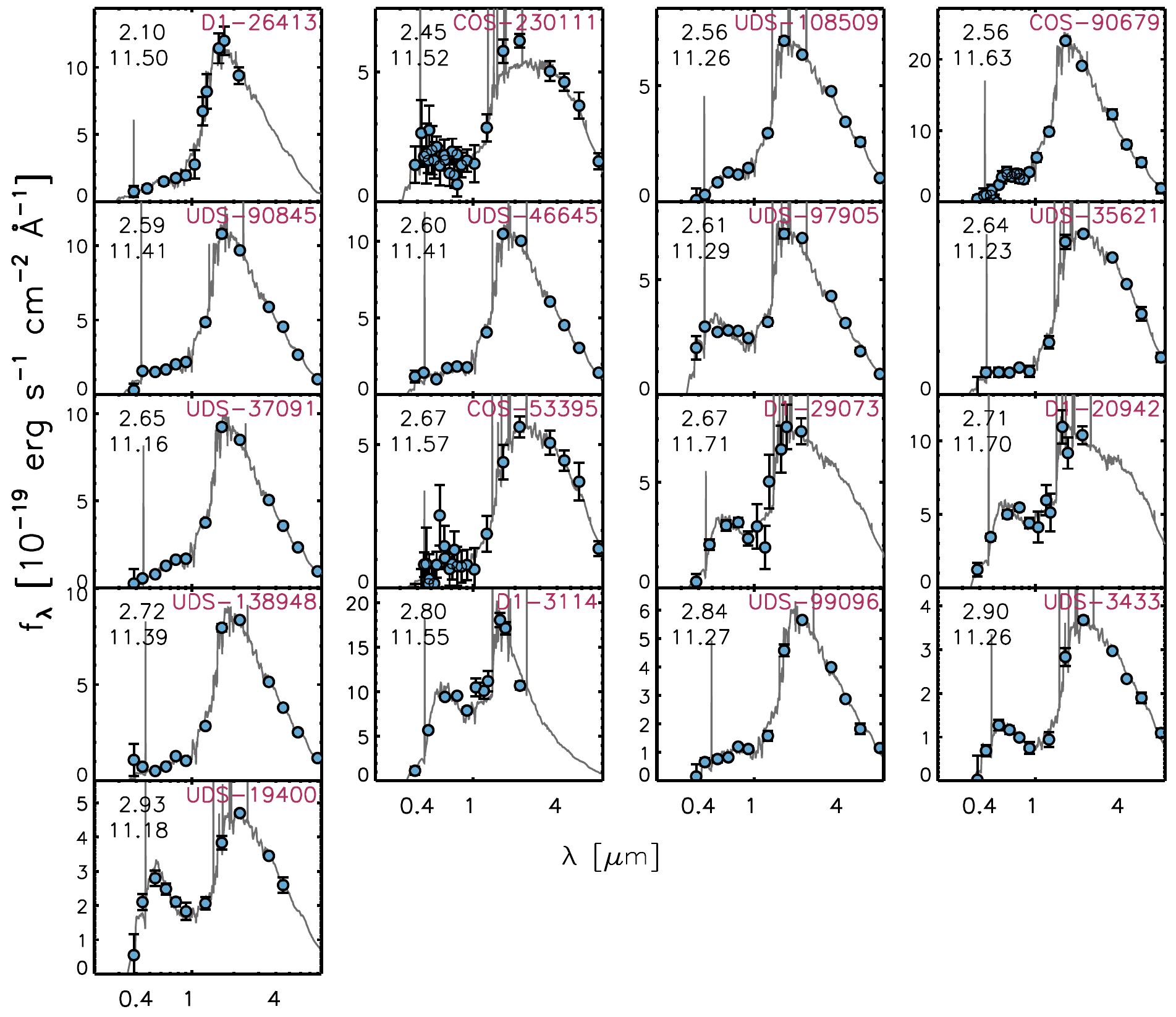}
\caption{The observed UV-IR spectral energy distributions for 17 of the 37 massive galaxies at $1.5<z<3$ selected for targeted \textit{HST/WFC3} follow-up observations. 
Photometry from parent catalogs are shown in blue filled symbols, in units of $10^{-19}$ erg~cm$^{-2}$~s$^{-1}$~\AA$^{-1}$
and the gray curves represent the best-fit EAZY templates. The ID, $z_{phot}$ and stellar mass ($\log(M_{*}/M_{\odot})$) of targets are listed in each panel, using the abbreviation D4 and CDF to denote targets in the COSMOS, CFHTD-1, CFHTD-4 and ECDFS fields, respectively.  } \label{fig-sed2}
\end{figure*}

Figures~\ref{fig-sed1} and \ref{fig-sed2} display the observed SEDs of this sample, with best-fit EAZY templates overplotted in gray. 
The SEDs are well sampled with the available medium- and broad-band photometry, and a strong rest-frame optical break is evident in all targets, indicative of relatively evolved stellar populations.

\section{Analysis} \label{sec3-analysis}

\begin{figure}
\epsscale{1}
\plotone{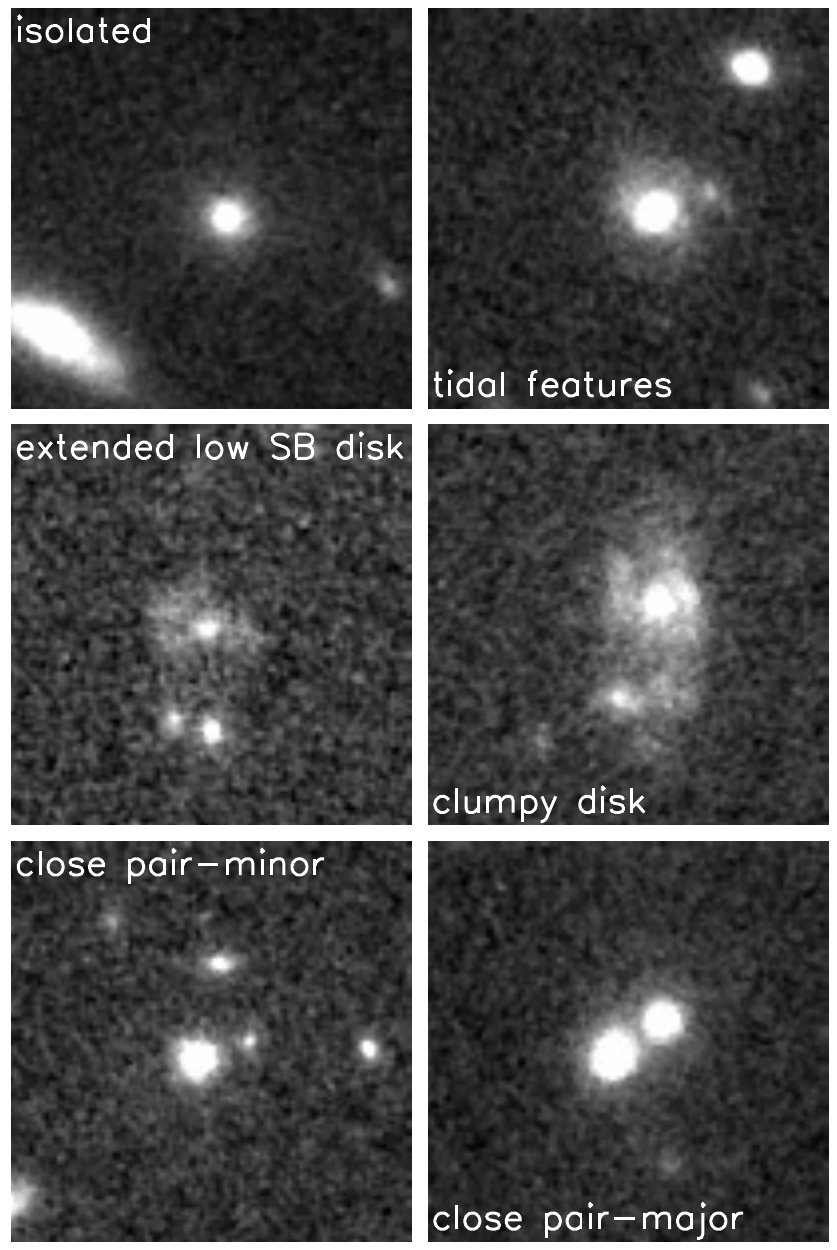}
\caption{$H_{160}$ stamps of targets displaying the structural diversity of the sample of very massive galaxies at $1.5<z<3.0$. Size of each image stamp is $6^{''}\times6^{''}$.} \label{fig-prodiv}
\end{figure}

Visually investigating the $H_{160}$ images of the 37 targets in this study reveals that very massive galaxies at $1.5<z<3.0$ are morphologically diverse, in contrast to their local universe counterparts. Figure~\ref{fig-prodiv} displays examples for the variety of structures observed: an isolated and morphologically undisturbed galaxy, a target with faint tidal features, a galaxy exhibiting the presence of an extended low-surface brightness disk, a galaxy displaying prominent features of disturbance and close-pair systems. 
Interestingly, 13 targets ($\sim 1/3$ of total sample) are found to be composed of multiple objects that are not resolved in the ground-based images (indicated with red stars in the corresponding $H_{160}$ panels of Figures \ref{fig3-galfit3}, \ref{fig3-galfit1} and \ref{fig3-galfit2}).
We further explore this effect of source multiplicity on the high-mass end of the galaxy stellar mass function (SMF) at $1.5<z<3$ inferred from ground-based NIR galaxy surveys in Section~\ref{sec3-smf}.

\subsection{Modeling 2-D Light Profiles} \label{sec3-modeling}

\begin{deluxetable*}{lccccccc}[!h]
\centering
\tabletypesize{\footnotesize}
\tablecaption{Best-fit GALFIT Structural Parameters}
\tablehead{ \colhead{Target} & \colhead{RA} & \colhead{DEC} &
	\colhead{$z_{\rm phot}$} & \colhead{$\log(M_{*}/M_{\odot})$}   & \colhead{$R_e$} & \colhead{n} & \colhead{$b/a$} \\
					& \colhead{(h:m:s)} & \colhead{(d:m:s)}	&       			  & \colhead{(dex)}					 & 	\colhead{(kpc)}   & 		&      }
\startdata
CFHTD1-2580$^{*}$ & $02:24:42.279$  & $-04:53:34.98$ & $1.51^{+0.07}_{-0.07}$  &  $11.38$  &  $6.61^{+1.20}_{-1.48}$  &  $7.42^{+0.54}_{-0.77}$  &  $0.32^{+0.02}_{-0.02}$ \\ 
CFHTD1-45042$^{*}$ & $02:24:19.551$    &  $-04:08:27.76 $ & $1.52^{+0.07}_{-0.07}$  &  $11.66$  &  $25.40^{+0.00}_{-0.00}$  &  $5.94^{+0.22}_{-0.21}$  &  $0.66^{+0.21}_{-0.05}$ \\ 
CFHTD4-32377 & $22:15:57.971$ & $-17:40:20.48 $ & $1.53^{+0.07}_{-0.06}$  &  $11.62$ &  $6.40^{+0.60}_{-0.60}$  &  $5.72^{+0.68}_{-0.68}$  &  $0.93^{+0.01}_{-0.01}$ \\ 
CFHTD4-38902 &  $22:15:56.557  $ & $-17:26:55.81$ & $1.58^{+0.11}_{-0.09}$  &  $11.61$  &  $3.42^{+1.02}_{-0.60}$  &  $4.29^{+1.35}_{-1.03}$  &  $0.59^{+0.02}_{-0.02}$ \\ 
CFHTD4-31678$^{*}$ & $ 22:15:44.443 $ & $ -17:41:50.14 $ & $1.59^{+0.08}_{-0.08}$  &  $11.27$  &  $3.21^{+0.08}_{-0.08}$  &  $3.78^{+0.33}_{-0.33}$  &  $0.91^{+0.12}_{-0.03}$ \\ 
ECDFS-712 & $ 03:31:46.077 $ & $  -28:00:26.48 $ &  $1.64^{+0.10}_{-0.10}$  &  $11.50$  &  $7.76^{+0.07}_{-0.07}$  &  $2.58^{+0.08}_{-0.08}$  &  $0.82^{+0.00}_{-0.00}$ \\ 
ECDFS-8431 & $ 03:32:42.636 $ &  $-27:38:15.93 $ &  $1.64^{+0.08}_{-0.09}$  &  $11.45$  &  $9.26^{+0.39}_{-0.39}$  &  $3.77^{+0.22}_{-0.22}$  &  $0.85^{+0.01}_{-0.01}$ \\ 
CFHTD1-24028$^{*}$ & $ 02:27:09.848 $ & $ -04:44:53.99 $ & $1.64^{+0.12}_{-0.11}$  &  $11.53$  &  $4.50^{+0.17}_{-0.17}$  &  $3.26^{+0.31}_{-0.31}$  &  $0.65^{+0.04}_{-0.04}$ \\ 
CFHTD1-7722 &  $ 02:24:09.991 $ & $ -04:46:7.83  $ &$1.66^{+0.14}_{-0.13}$  &  $11.50$  &  $5.80^{+0.13}_{-0.13}$  &  $2.35^{+0.13}_{-0.12}$  &  $0.89^{+0.01}_{-0.01}$ \\ 
ECDFS-36 & $03:31:54.522 $ & $  -28:02:22.66$  & $1.74^{+0.08}_{-0.08}$  &  $11.41$  &  $10.27^{+0.00}_{-0.00}$  &  $0.42^{+0.00}_{-0.00}$  &  $0.49^{+0.00}_{-0.00}$ \\ 
COSMOS-75358$^{*}$ & $10:02:28.491 $ & $ 02:02:13.70 $ &$1.79^{+0.04}_{-0.04}$  &  $11.50$  &  $3.36^{+0.23}_{-0.20}$  &  $3.68^{+0.35}_{-0.36}$  &  $0.73^{+0.17}_{-0.08}$ \\ 
COSMOS-97596$^{*}$ & $10:02:32.428 $ & $ 02:22:40.59 $ & $1.84^{+0.13}_{-0.09}$  &  $11.28$  &  $7.55^{+1.26}_{-1.29}$  &  $3.46^{+0.45}_{-0.58}$  &  $0.46^{+0.10}_{-0.03}$ \\ 
COSMOS-189978 &  $10:02:14.418  $ & $02:35:11.92$  & $1.84^{+0.16}_{-0.14}$  &  $11.50$  &  $14.92^{+3.32}_{-3.74}$  &  $3.32^{+0.27}_{-0.30}$  &  $0.93^{+0.04}_{-0.03}$ \\ 
ECDFS-4444$^{*}$ &  $03:33:11.482 $ & $ -27:49:16.06$  & $1.87^{+0.16}_{-0.15}$  &  $11.43$  &  $3.25^{+0.10}_{-0.11}$  &  $4.82^{+0.36}_{-0.36}$  &  $0.84^{+0.06}_{-0.05}$ \\ 
COSMOS-207160$^{*}$ & $10:00:33.479 $ & $ 02:28:54.74$ & $1.90^{+0.13}_{-0.10}$  &  $11.25$  &  $6.71^{+1.93}_{-1.88}$  &  $8.00^{+0.00}_{-0.00}$  &  $0.58^{+0.24}_{-0.03}$ \\ 
COSMOS-51726$^{*}$ & $ 09:59:23.943 $ & $ 01:44:11.52$ & $1.94^{+0.05}_{-0.05}$  &  $11.46$  &  $4.62^{+0.04}_{-0.04}$  &  $2.64^{+0.12}_{-0.12}$  &  $0.92^{+0.27}_{-0.02}$ \\ 
COSMOS-37208 & $09:59:42.594 $ & $ 01:55:01.55 $ & $1.95^{+0.07}_{-0.07}$  &  $11.54$  &  $6.35Ÿ^{+0.01}_{-0.01}$  &  $1.51^{+0.02}_{-0.02}$  &  $0.77^{+0.00}_{-0.00}$ \\ 
CFHTD4-39098 & $ 22:16:52.676 $ & $ -17:26:29.17$ &   $1.99^{+0.15}_{-0.14}$  &  $11.55$  &  $3.95^{+0.06}_{-0.07}$  &  $4.09^{+0.33}_{-0.33}$  &  $0.89^{+0.01}_{-0.01}$ \\ 
COSMOS-71932 &  $10:01:40.598 $ & $ 01:58:57.47$ & $2.07^{+0.10}_{-0.11}$  &  $11.70$  &  $4.17^{+1.63}_{-1.06}$  &  $4.67^{+1.49}_{-1.14}$  &  $0.75^{+0.02}_{-0.02}$ \\ 
COSMOS-103240 & $10:00:47.179 $ & $ 01:59:19.56$  & $2.09^{+0.09}_{-0.08}$  &  $11.49$  &  $28.0^{+4.19}_{-2.55}$  &  $8.00^{+0.00}_{-0.00}$  &  $0.60^{+0.01}_{-0.01}$ \\ 
CFHTD1-26413 & $2:26:42.090 $ &  $ -04:40:39.88 $ & $2.10^{+0.21}_{-0.21}$  &  $11.50$  &  $3.12^{+0.04}_{-0.04}$  &  $2.65^{+0.17}_{-0.17}$  &  $0.65^{+0.01}_{-0.01}$ \\ 
COSMOS-230111 & $10:01:23.525 $ & $ 02:45:40.07 $ & $2.45^{+0.17}_{-0.17}$  &  $11.52$  &  $10.53^{+15.5}_{-9.21}$  &  $2.72^{+2.14}_{-2.62}$  &  $0.75^{+0.10}_{-0.18}$ \\ 
UDS-108509 &  $02:18:46.503 $ & $ -04:59:29.3$&  $2.56^{+0.08}_{-0.08}$  &  $11.26$  &  $3.42^{+0.02}_{-0.02}$  &  $1.73^{+0.06}_{-0.06}$  &  $0.64^{+0.01}_{-0.01}$ \\ 
COSMOS-90679 & $10:01:57.001$ & $  02:16:12.14$ & $2.56^{+0.07}_{-0.07}$  &  $11.63$  &  $6.27^{+0.15}_{-0.15}$  &  $4.91^{+0.34}_{-0.34}$  &  $0.67^{+0.01}_{-0.01}$ \\ 
UDS-90845 &   $ 02:17:12.786  $ & $-05:04:49.97 $ & $2.59^{+0.08}_{-0.08}$  &  $11.41$  &  $8.40^{+2.51}_{-1.46}$  &  $8.00^{+0.00}_{-0.00}$  &  $0.82^{+0.02}_{-0.02}$ \\ 
UDS-46645 &  $ 02:16:59.092  $ & $ -05:18:07.07 $ &  $2.60^{+0.08}_{-0.08}$  &  $11.41$  &  $1.98^{+0.07}_{-0.07}$  &  $3.08^{+0.23}_{-0.23}$  &  $0.52^{+0.03}_{-0.03}$ \\ 
UDS-97905 &  $ 02:16:08.893 $ & $ -05:02:37.69 $ &  $2.61^{+0.06}_{-0.06}$  &  $11.29$  &  $5.25^{+0.51}_{-0.61}$  &  $7.28^{+0.78}_{-0.82}$  &  $0.66^{+0.01}_{-0.01}$ \\ 
UDS-35621$^{*}$ &   $ 02:18:19.404  $ & $-05:21:32.67 $ & $2.64^{+0.11}_{-0.10}$  &  $11.23$  &  $6.14^{+1.61}_{-2.73}$  &  $5.94^{+0.98}_{-1.64}$  &  $0.74^{+0.17}_{-0.20}$ \\ 
UDS-37091$^{*}$ &   $ 02:17:09.297$  & $ -05:21:07.42 $ & $2.65^{+0.09}_{-0.08}$  &  $11.16$  &  $1.60^{+0.07}_{-0.07}$  &  $4.03^{+0.54}_{-0.54}$  &  $0.43^{+0.31}_{-0.01}$ \\ 
COSMOS-53395 & $09:58:10.638 $ & $ 01:45:31.92 $  &$2.67^{+0.25}_{-0.26}$  &  $11.57$  &  $4.34^{+0.14}_{-0.14}$  &  $0.69^{+0.03}_{-0.03}$  &  $0.77^{+0.01}_{-0.01}$ \\ 
CFHTD1-29073 &  $02:26:25.408  $ & $ -04:36:5.42$ & $2.67^{+0.20}_{-0.19}$  &  $11.71$  &  $7.01^{+0.25}_{-0.25}$  &  $4.91^{+0.32}_{-0.32}$  &  $0.83^{+0.01}_{-0.01}$ \\ 
CFHTD1-20942$^{*}$ & $02:26:49.300$ & $ -04:49:20.53 $ & $2.71^{+0.11}_{-0.11}$  &  $11.70$  &  $1.61^{+0.05}_{-0.05}$  &  $2.40^{+0.23}_{-0.23}$  &  $0.89^{+0.12}_{-0.05}$ \\ 
UDS-138948 &   $ 02:17:34.679 $ & $ -04:50:09.92 $ &  $2.72^{+0.10}_{-0.10}$  &  $11.39$  &  $4.76^{+0.13}_{-0.13}$  &  $2.81^{+0.22}_{-0.22}$  &  $0.72^{+0.01}_{-0.01}$ \\ 
CFHTD1-3114 & $02:24:29.921 $ & $-04:52:38.92$ & $2.80^{+0.08}_{-0.08}$  &  $11.55$  &  $1.36^{+0.08}_{-0.08}$  &  $7.42^{+1.00}_{-1.00}$  &  $0.63^{+0.04}_{-0.04}$ \\ 
UDS-99096 &   $ 02:17:09.861 $ & $ -05:02:17.17 $ & $2.84^{+0.14}_{-0.12}$  &  $11.27$  &  $9.13^{+7.4}_{-3.8}$  &  $8.00^{+0.00}_{-0.00}$  &  $0.96^{+0.03}_{-0.04}$ \\ 
UDS-3433 &  $ 02:17:56.693 $ & $ -05:31:16.65 $ &  $2.90^{+0.15}_{-0.14}$  &  $11.26$  &  $32.5^{+11.7}_{-16.1}$  &  $6.54^{+0.77}_{-1.39}$  &  $0.70^{+0.01}_{-0.01}$ \\ 
UDS-19400$^{*}$ & $ 02:19:11.929 $ & $ -05:26:24.62 $ & $2.93^{+0.17}_{-0.21}$  &  $11.18$  &  $3.70^{+0.24}_{-0.27}$  &  $3.78^{+0.78}_{-0.79}$  &  $0.31^{+0.55}_{-0.04}$ \\
\enddata
\tablecomments{Properties of the $1.5<z<3$ massive galaxy sample selected for targeted imaging follow-up. Listed photometric redshifts are $z_{peak}$ values from parent catalogs. ID's with $^{*}$ indicate targets that are resolved as multiple components/close pairs in HST $H_{160}$ imaging. The listed stellar masses correspond to values from the parent catalogs for targets resolved as single objects in both ground and HST imaging, whereas for the targets resolved as multiple components it corresponds to the decomposed (catalog) stellar masses calculated in Section~\ref{sec3-decomp}. Also listed are the best-fit GALFIT structural parameters ($R_{e}$, S\'ersic index $n$ and axis ratio $b/a$).} \label{tab-galfitpar}
\end{deluxetable*}

The \textit{HST} $H_{160}$ data were reduced with AstroDrizzle in a similar manner as CANDELS imaging \citep{koekemoer11}. The exposures from the 4-point dither pattern were combined to a final pixel scale of $0.^{''}06$. The total exposure time for each target ranged from $\sim500-2400$~s. 

We used GALFIT \citep{peng10} to model the two-dimensional light profiles and to obtain the structural parameters for the $\log(M_{*}/M_{\odot}$)$>11.25$ sample of galaxies at $1.5<z<3$ using cut-outs of size $12^{''}\times12^{''}$ centered around each target's position. 

An empirical point spread function (PSF) to be included in GALFIT modeling was created for each target by median-stacking the sky-subtracted, two-dimensional light profiles of bright, unsaturated stars located within its frame. The \textit{HST} imaging resolution (FWHM$\approx 0.18^{''}$) is a factor of $\approx4-5$ greater compared to the seeing of the ground-based NIR observations ($\sim0.^{''}8$). 
 When extended, bright objects were present in image stamps, we created a bad pixel map to mask out these components when fitting with GALFIT. Following \citet{vanderwel14a} and others, the sky background level was kept fixed in the fitting procedure, which we estimated as the mode of sky pixel values after masking out all objects in the image stamps. 
We also repeated this analysis by allowing GALFIT to fit for the sky background, finding that the results were quantitatively robust against the specific treatment for the sky background. We constructed uncertainty maps to be used as inputs for GALFIT by adding the Poisson noise across the images and the noise calculated from the inverse variance maps produced by AstroDrizzle (corresponding to the instrumental noise) in quadrature. 

A single-component S\'ersic model was used to model the light profiles of the targets simultaneously with all other objects located within their $H_{160}$ stamps. 
Specifically, we used GALFIT to determine the best-fit total magnitude ($H_{160}$), half-light radius along the the semi-major axis ($r_{\rm 1/2,maj}$), S\'ersic index ($n$), axis ratio ($b/a$), position angle (PA) and the centroid for each object.
A constraint file was created to force GALFIT to restrict the fit S\'ersic indices between $n=0.25-8$, the semi-major half-light radii between 1-50 pixels (50 pixels = 3$^{''}$, corresponding to $\sim 23-25$~kpc at $z\approx 1-3$) and the total magnitudes of sources within $\pm3$~mag of the parent catalog \textit{H} band photometry. When the radius along the semi-major axis reached the extreme value of $r_{\rm 1/2,maj}=50$~pixels, we reran GALFIT after relaxing the upper constraint to  $r_{\rm 1/2,maj}=100$~pixels. 

We initially modeled the observed light profiles multiple times for each target by varying initial guesses to obtain a measure of the dependence of the best-fit S\'ersic parameters on GALFIT inputs. 
Specifically, we ran GALFIT 100 times for each target by selecting the initial input values for the effective radius and $n$ from a random distribution of values between 1-20 pixels and 1-6, respectively.
When estimating confidence limits, we only considered the GALFIT models that yielded valid results, discarding models that did not converge numerically. The confidence limit for each structural parameter was determined by using the $1\sigma$ standard deviation of its distribution (i.e., by integrating the probability distribution of each parameter from the extremes until the integrated probability is equal to $0.3173/2 = 0.1586$.)

As galaxies are more compact at high-$z$, we additionally investigated the effect of PSF model choice on the estimated structural parameters by running GALFIT on each target using all the empirical PSFs (36 additional for each target). 
The final $1\sigma$ confidence limits were then calculated by combining the scatter of S\'ersic parameters derived in this manner with the former $1\sigma$ distribution values in quadrature. 

Table~\ref{tab-galfitpar} lists the photometric redshifts, and the photometrically derived stellar masses (after the decomposing for the blended sources, indicated with $^*$; see Section~\ref{sec3-decomp}) of the targeted sample, along with their best-fit GALFIT structural properties and corresponding 1$\sigma$ uncertainties.

\begin{figure*}
\plotone{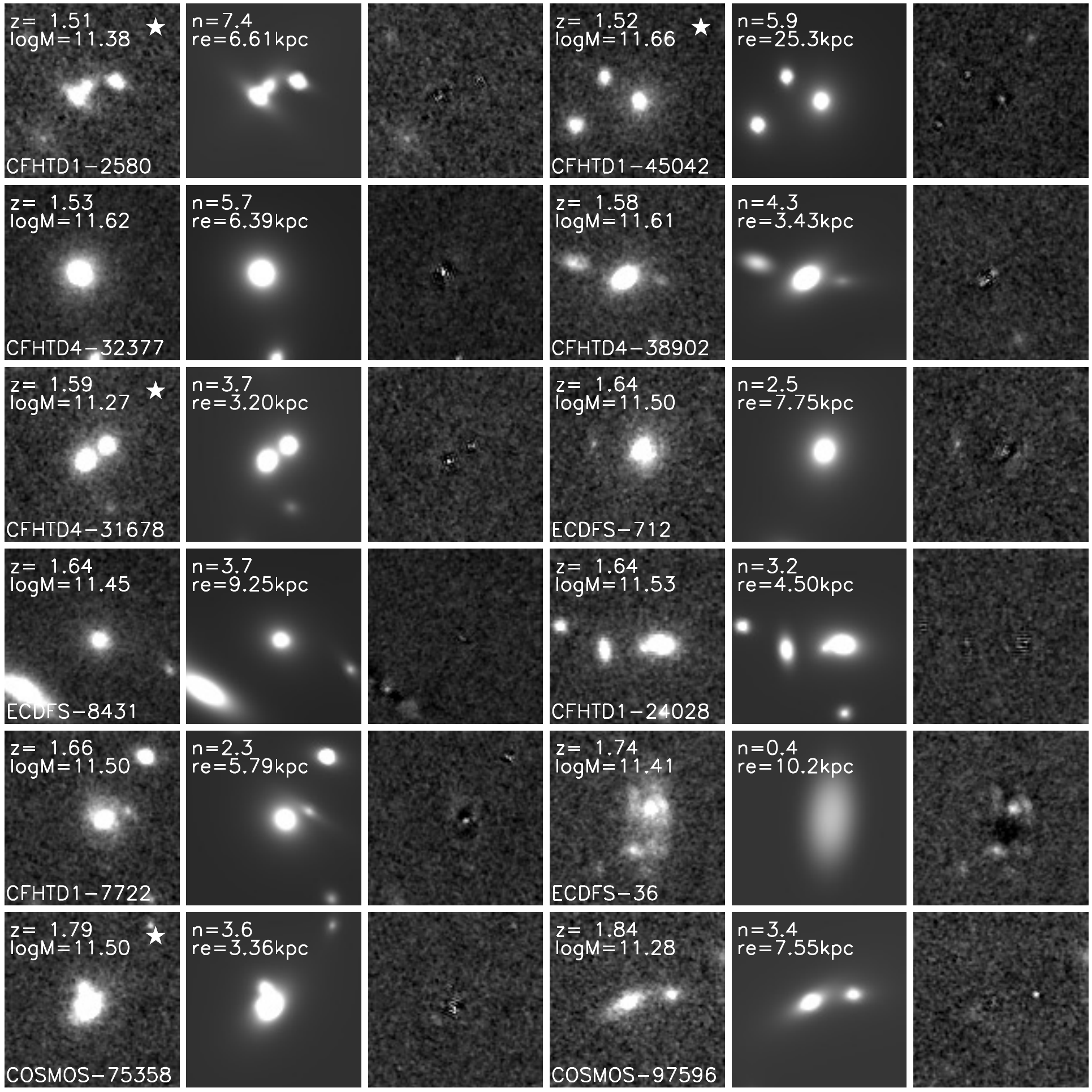}
\caption{The variety of the 2D light profiles of the $1.5<z<3.0$ massive galaxies. The \textit{HST} $H_{160}$ image cutouts (panels with target name, redshift, and stellar mass in legend), best-fit 2D light profile (panels with $n$ and $R_{e}$ indicated) and the residual image is displayed for each target. Panels with white stars indicate targets that are resolved as multiple components in the \textit{HST} $H_{160}$ imaging.
Panels are $6^{''}$ on each side.}  \label{fig3-galfit3}
\end{figure*}

\begin{figure*}
\centering
\epsscale{1}

\plotone{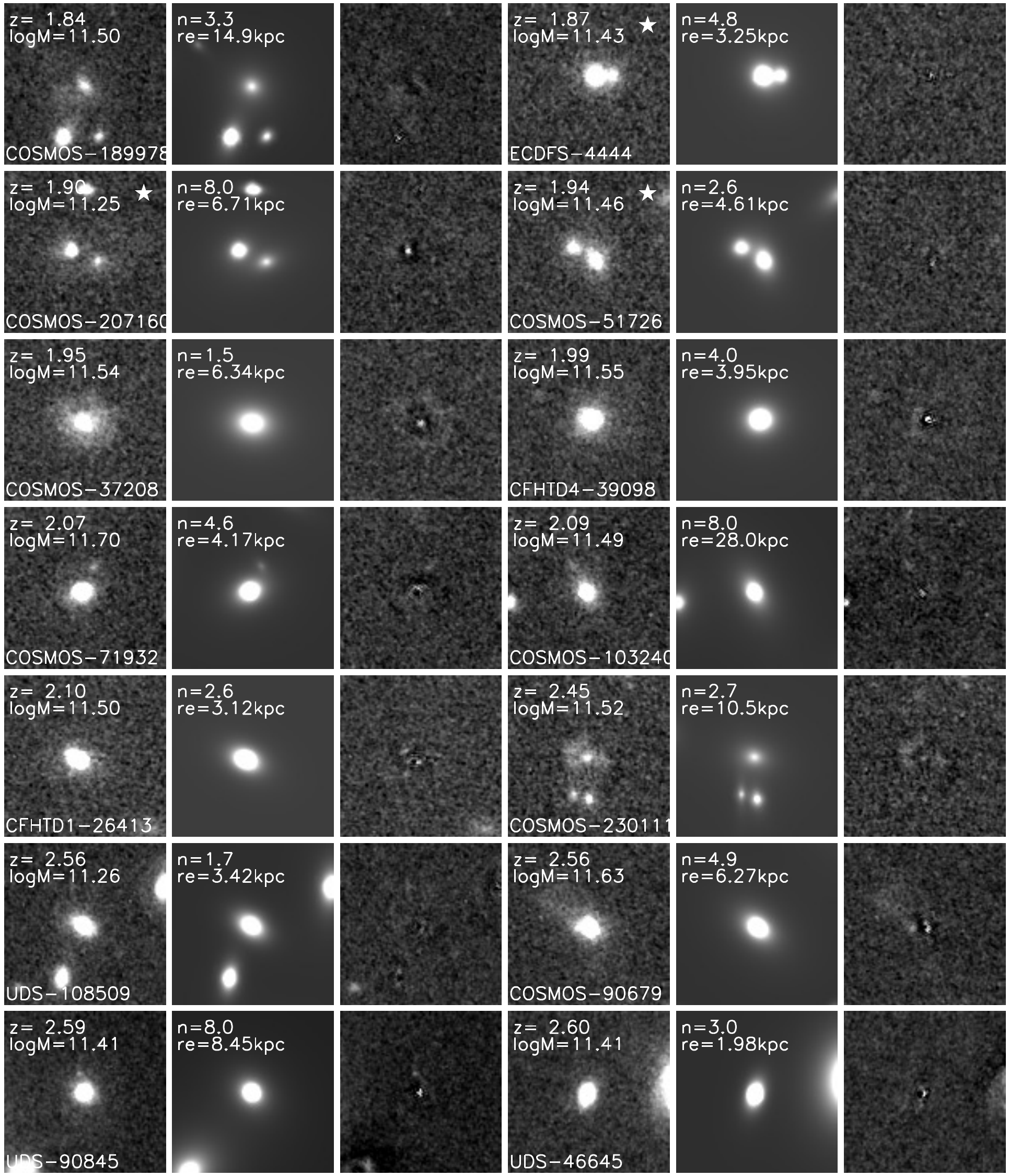}

\caption{See caption for Figure~\ref{fig3-galfit3}
\label{fig3-galfit1}}
\end{figure*}

\begin{figure*}
\plotone{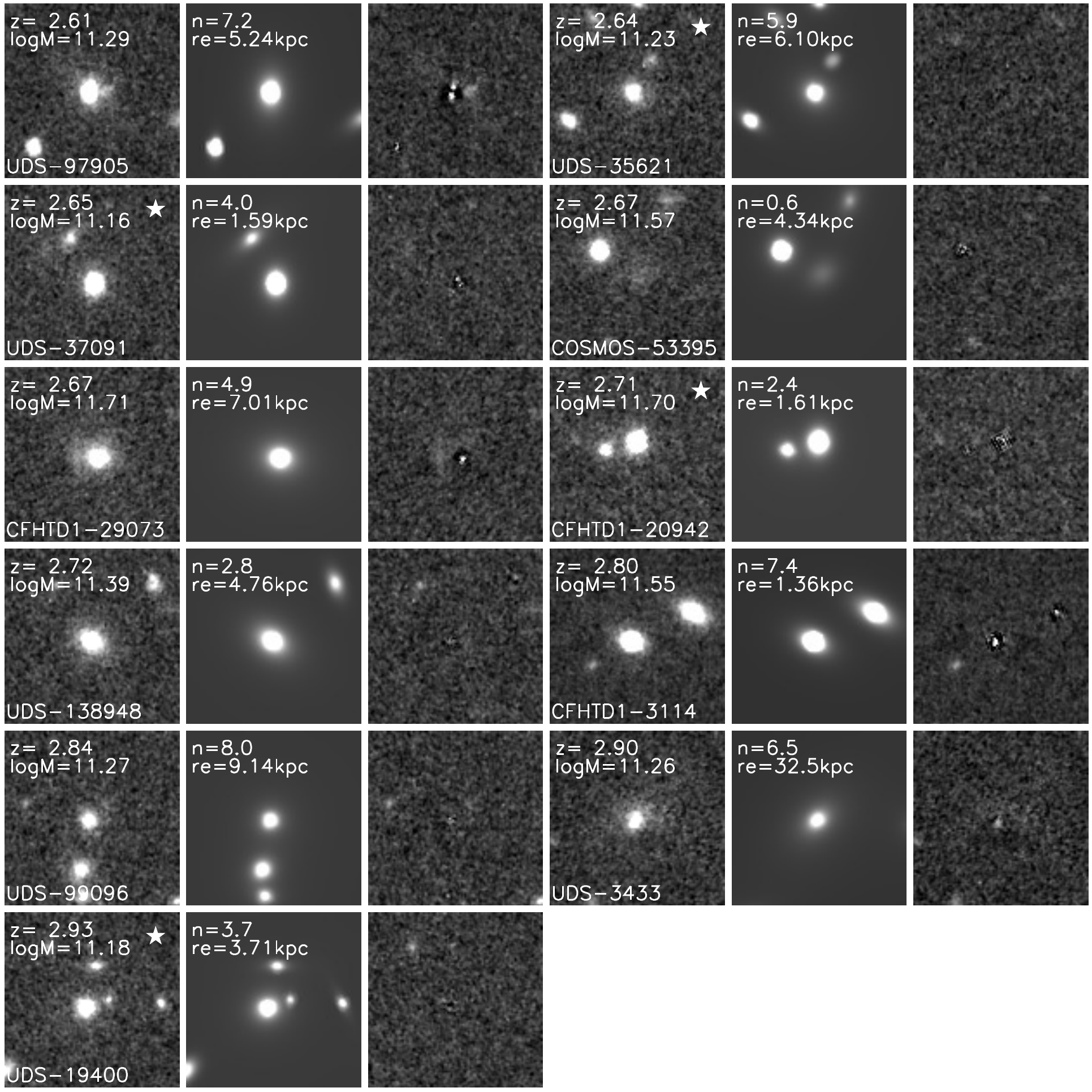}
\caption{See caption for Figure~\ref{fig3-galfit3}
\label{fig3-galfit2}}
\end{figure*}

Figures \ref{fig3-galfit3}, \ref{fig3-galfit1} and \ref{fig3-galfit2} show the GALFIT-modeled $H_{160}$ stamps, along with the best-fit GALFIT 2D models and the residuals (displayed using identical scaling in the panels for each target). The legend of the direct $H_{160}$ imaging panels lists the target ID, $z_{phot}$ and stellar mass, whereas the best-fit structural parameters are listed in the legends of the GALFIT model panels. Panels with red stars indicate the targets that are revealed to be multiple sources in the $H_{160}$ imaging.

\subsection{Stellar Mass Decomposition of HST-Resolved Close Galaxy Pairs}\label{sec3-decomp}

In this section we focus on the targets that are resolved as multi-component systems in $H_{160}$ imaging. 
We used the observed $H_{160}$ magnitudes of close galaxy pairs (centrals and companions) as proxies to decompose the stellar masses of ground-based blended objects. While this method is not ideal, we will show that it is an appropriate first-order approximation to assume that the central and companion galaxies have stellar masses proportional to their light observed in the $H_{160}$ band. In other words, we assume identical mass to light ratios 
($M_{*}/L_{H}$) for the HST resolved components and use their observed $H_{160}$ band fluxes as direct tracers for their underlying stellar masses. This inherently brings with it two additional assumptions for the properties of \textit{HST} resolved close pairs: 1) that the close pairs are physically associated -- i.e., not chance superpositions of objects at different redshifts along the line of sight, and 2) that the central and companion galaxies have similar stellar populations. Strict proof for the validity of these assumptions requires spectroscopic redshift identifications of the resolved components, and multi-wavelength, space-based imaging of all targeted objects, which are currently not available.

To address the first assumption, we checked the publicly available spectroscopic catalogs for these widely studied fields (VIMOS Ultra Deep Survey, \citealt{lefevre15, tasca17}; zCOSMOS, \citealt{lilly07}; DEIMOS 10k spectroscopic catalog of the COSMOS field, \citealt{hasinger18}; VANDELS spectroscopic survey of the UDS and CDFS fields, \citealt{mclure18}), finding no matches. Only one of the targets (COSMOS-207160 - that has two resolved component centers located $\sim 1.1^{''}$ apart) was identified in the grism redshift catalog of 3D-HST \citep{momcheva16}. This target has a $z_{\rm peak}=1.91^{+0.13}_{-0.11}$ in the parent catalog from which it was selected (UVISTA). The $z_{\rm grism}$ in the 3D-HST redshift catalog for the resolved sources are $2.05^{+0.01}_{-0.02}$ and $2.36^{+0.0001}_{-0.02}$. 
There are no discernible color differences between the two resolved sources. Although the true physical pairs cannot be identified without spectroscopic redshifts, at such small angular scales, it is more likely that the pairs are physically associated rather than chance alignment \citep{quadri12}.

\begin{figure}
\epsscale{1.0}
\plotone{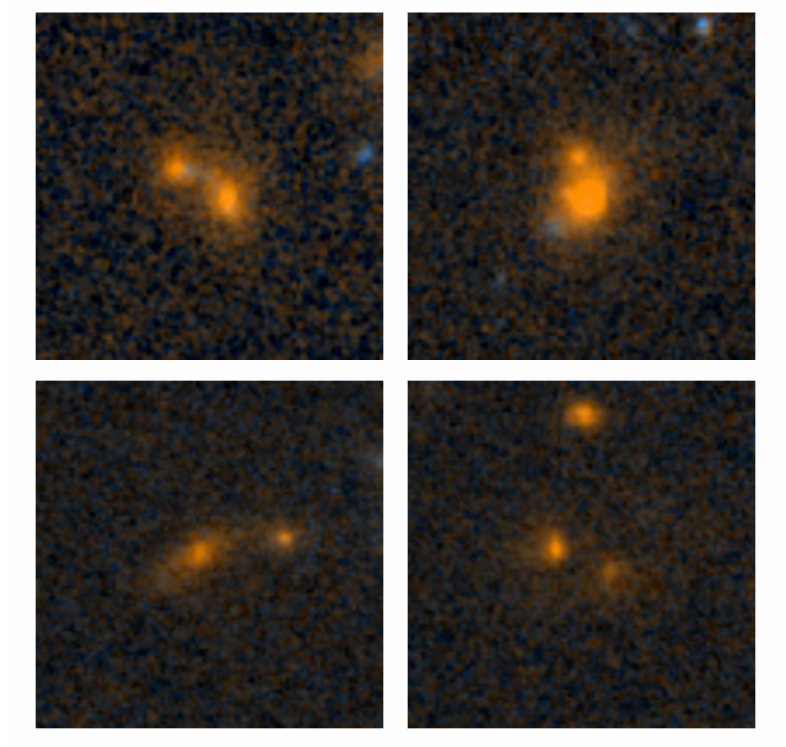}
\caption{Color images for the four targets in our sample located in the COSMOS field that have publicly available ACS imaging.} \label{fig3-RGB}
\end{figure}

In order to investigate the validity of the second assumption, i.e. that the $M_{*}/L_{H}$ of close pairs are similar, we utilized the publicly available deep \textit{HST} ACS/F814W band (hereafter, $i_{814}$) imaging of the COSMOS field to investigate the color differences between resolved sources, as a proxy of different stellar populations. Figure~\ref{fig3-RGB} shows the $i_{814}$+$H_{160}$ color composite images for the four targets with available $i_{814}$ imaging. Visually, there are no discernible differences between the colors of resolved objects, supporting the scenario that they do not have significantly different stellar populations. 
To quantitatively assess the color differences of the \textit{HST} resolved components, we calculated the $i_{814}$ and $H_{160}$ magnitudes using a circularized aperture of $d=0.^{''}3$ centered at their locations. Not surprisingly, due to the faintness of these targets just below $\lambda_{obs}\sim1\mu$m (Figure~\ref{fig-sed1}), they are barely detected/resolved in $i_{814}$ imaging at best, and therefore their calculated magnitudes have significant 
uncertainties associated with them due to Poisson statistics (the uncertainty in the calculated color differences is dominated by this term). We estimated the noise due to variations in the sky background by calculating the $\sigma$ of the Gaussian fit to the distribution of fluxes measured in $d=0.^{''}3$ apertures on empty regions of the sky, and added this value in quadrature to the uncertainty on the measured $i_{814}$ magnitudes. 
The color differences between resolved close pairs range from  $\Delta$($i_{814}-H_{160})=0.3 - 2.3$~mag, however, the colors of pairs are all consistent with each other within 1$\sigma$ uncertainties. 

In addition to investigating the size-stellar mass relation at its extreme massive end, we are seeking to constrain the effect of blending in ground-based surveys on the inferred number density of very massive galaxies at $z>1.5$. Assuming similar $M_{*}/L_{H}$ ratios for the blended objects translates to constraining the maximum allowed change due to blending on the extreme massive end of the SMF at $z>1.5$. In fact, even if
there are color differences that we are not able to discern, the assumption that resolved close pairs have identical $M_{*}/L_{H}$ ratios \textit{maximally} reduces the stellar mass of the central (brighter, and hence more massive) galaxy. If the fainter companions have younger stellar populations, it is expected that $(M_{*}/L_{H})_{companion}<(M_{*}/L_{H})_{central}$, which would work to decrease the stellar mass allocated to the companion. Hence, assuming identical $M_{*}/L_{H}$ ratios for HST resolved close pairs of galaxies sets a conservative lower limit to the stellar mass of the central galaxies. 

\begin{figure}
\plotone{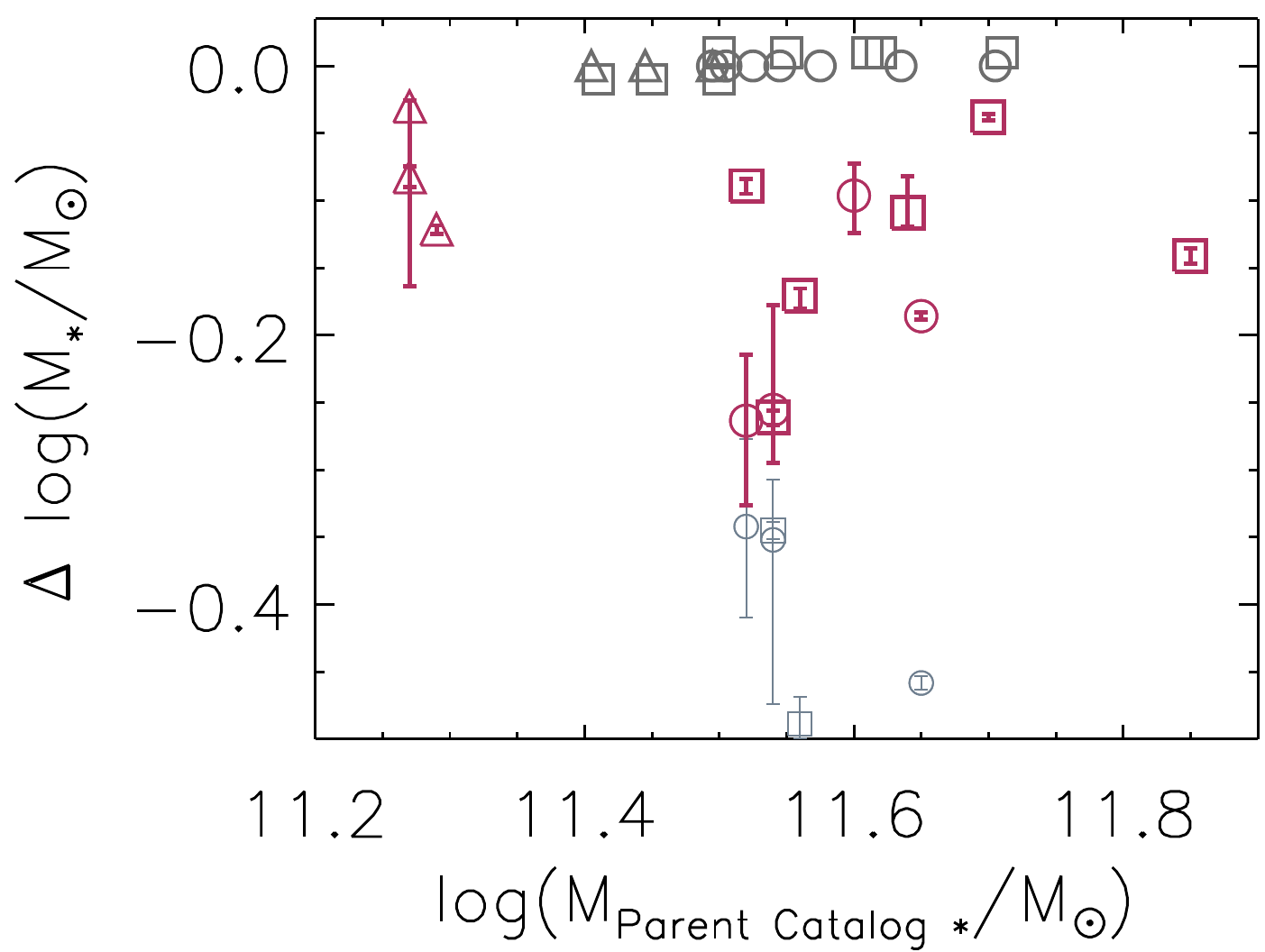}
\caption{The difference in stellar mass estimates once accounting for blending due to close-pairs. The brighter, more massive component of the blended targets are plotted in red, whereas the companions, i.e., the lower mass components are plotted in light gray. The stellar masses of the targets identified as being composed of close pairs are calculated by scaling the parent catalog stellar mass by the relative fluxes of the resolved components (see Section~\ref{sec3-decomp}). Targets selected from UltraVISTA, NMBS-II and UDS catalogs are represented as circles, squares and triangles.}
\label{fig3-massdecomp}
\end{figure}

Figure~\ref{fig3-massdecomp} serves to illustrate the effect of decomposing the stellar masses of resolved targets using this approach. The decomposed masses of $K$ band blended targets are shown with red and light gray symbols. The error bars on the deblended masses of close pairs are calculated using the $1\sigma$ standard deviation of their distribution in $H_{160}$ magnitude differences (i.e., $H_{160,{\rm central}}- H_{160, {\rm companion}}$ ). We caution that the stellar masses inferred for the less bright companions should not be taken at face value, rather, they are plotted in this figure to guide the eye to reflect the extent of blending. The median (mean) difference in the inferred stellar mass of the main/central galaxies is $\Delta\log(M_{*}) \sim 0.12 (0.14)$ dex. The difference in stellar mass inferred for the most major blends is $\Delta\log(M_{*}) \approx 0.25$ dex.

\section{Results} \label{sec3-results}

\subsection{The size-mass relation} \label{sec3-sizemass}

We used the radius along the semi-major axis of the half-light ellipse ($r_{\rm 1/2,maj}$) as a proxy for the sizes of our targets, rather than the often-calculated circularized effective radius in order to compare our results directly with \citet{vanderwel14a}. 
We converted the sizes of the modeled galaxies to the rest-frame 5000{\AA}, using Equations 1 and 2 in \citet{vanderwel14a} to correct for stellar mass and redshift dependent color gradients.  Table~\ref{tab-galfitpar} lists all sizes standardized to the rest-frame $\lambda=5000${\AA}.

\begin{figure*}
\epsscale{1}
\plotone{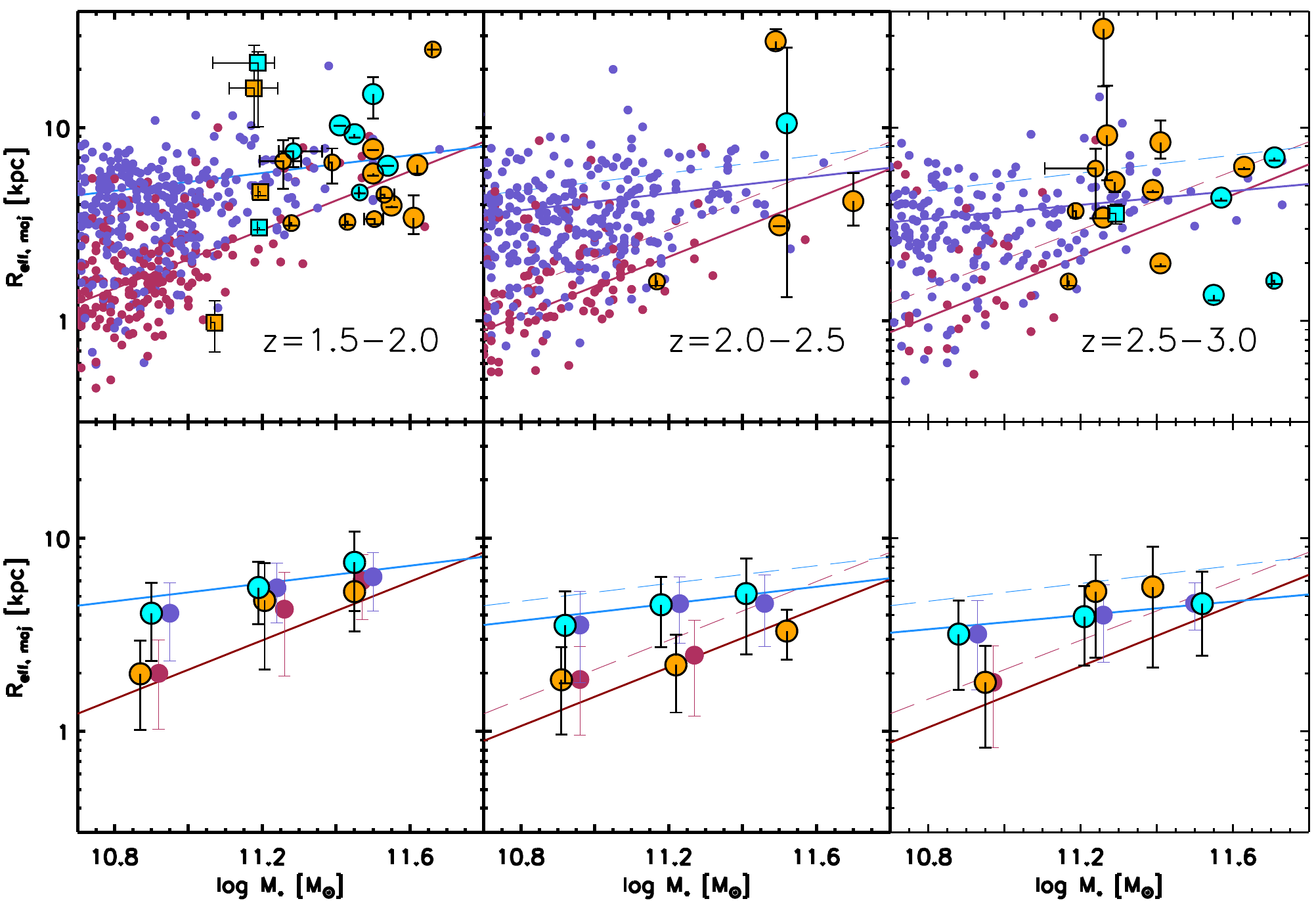}
\caption{Size - stellar mass relation in three redshift ranges, namely $1.5<z<2.0$, $2.0<z<2.5$, and $2.5<z<3.0$. Filled orange and cyan circles represent our targeted sample of quiescent and star-forming galaxies, respectively. The smaller red and blue filled circles represents the quiescent and star-forming galaxies from \citet{vanderwel14a} using the CANDELS survey. Top panels show individual size measurements. The bottom panels show the biweighted mean sizes and dispersions for the \citet{vanderwel14a} sample (red and purple), and including our sample (orange and cyan).  Also plotted are the size - stellar mass relations from \citet[red and blue solid lines]{vanderwel14a}.   
The $1.5<z<2.0$ relations are over-plotted in the higher redshift panels as dashed curves.}
\label{fig3-masssize}
\end{figure*}

Figure~\ref{fig3-masssize} shows our targeted sample of very massive galaxies at $1.5<z<3.0$ in the size-stellar mass diagram, along with the measurements from \citet{vanderwel14a}. The top panels show the size measurements for individual galaxies. The small red and blue points indicate the quiescent and  star-forming galaxies from the CANDELS sample from \citet{vanderwel14a} at the targeted redshifts.
Filled orange and cyan symbols represent our targeted sample of quiescent and star-forming galaxies, respectively. The filled circles indicate targets resolved both in the $K$ and $H_{160}$ bands (larger filled circles), or the central/main galaxies in blends (smaller filled circles). The square symbols represent the fainter companion galaxies with inferred deblended stellar mass $\log(M_{*}/M_{\odot})>11.0$. We note the large range in sizes observed at the extreme massive end probed by this sample. With this targeted \textit{HST} sample, the number of galaxies with robust size determinations increases by a factor of $\sim2$ in the lowest redshift bin, $1.5<z<2.0$, for both quiescent and star-forming galaxies with $\log(M_{*}/M_{\odot})>11.4$. Additionally, where the \citet{vanderwel14a} sample has only a single $\log(M_{*}/M_{\odot})>11.4$ (11.2) quiescent galaxy at $2.0<z<2.5$ ($2.5<z<3.0$), this sample adds crucial observations where CANDELS cannot probe due to its relatively narrow effective area. 

 The bottom panels of Figure~\ref{fig3-masssize} display the biweighted mean sizes inferred for massive $1.5<z<3.0$ galaxies. The size - stellar mass relations from \citet[red and blue solid lines]{vanderwel14a} for the corresponding redshift bins are plotted in each panel to aid the eye. We find that at $1.5<z<2.5$, the sizes of both star-forming and quiescent galaxies with $\log(M_{*}/M_{\odot})>11.2$ are relatively consistent with those found in \citet{vanderwel14a}. At $2.5<z<3.0$, the sizes of the very massive star-forming galaxies ($\log(M_{*}/M_{\odot})>11.4$) appears to follow the extrapolation of the lower stellar-mass galaxy sizes. Interestingly, the mean sizes for quiescent galaxies at $\log(M_{*}/M_{\odot})>11.2$ appear to be systematically larger than what is expected based on the extrapolation of the relation derived from lower stellar mass galaxies, hinting to either a steeper size - stellar mass relation of quiescent galaxies, or at a break at $\log(M_{*}/M_{\odot})\sim11.2$, such that more massive galaxies at $2.5<z<3$ have already reached their sizes, while the lower mass galaxies have yet to grow (see \citealt{patel17}).

\subsection{Effect of blending on the massive end of SMF at $1.5<z<3$} \label{sec3-smf}

\begin{figure*}
\epsscale{0.9}
\plotone{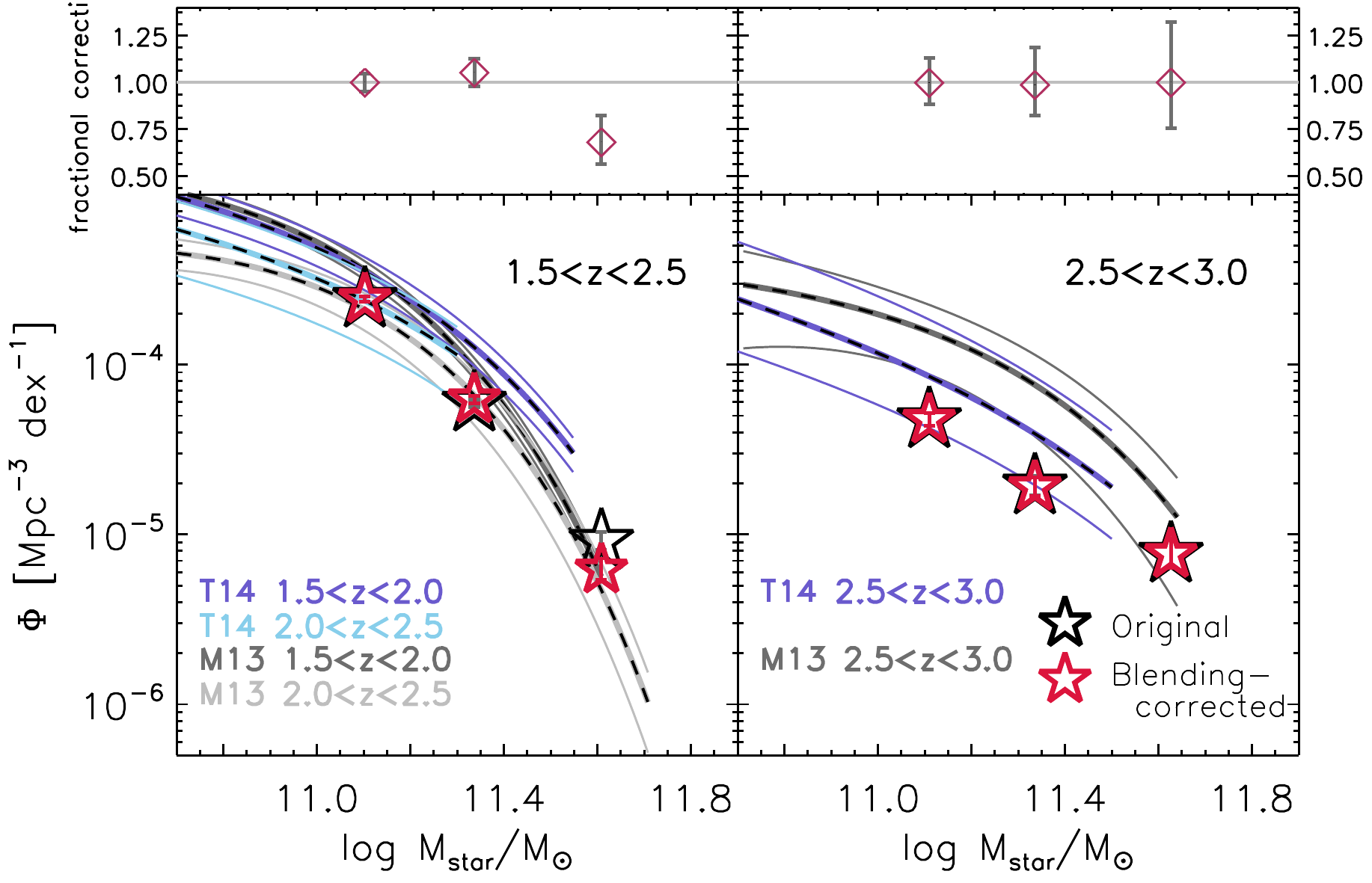}
\caption{\textit{Lower panels:} The stellar mass function, in stellar mass bins of $\Delta M_{*}=0.25$~dex before (black stars) and after correcting for the effect of blended galaxies in the ground-based $K$~band imaging (red stars). Also plotted are SMFs (thick curves with black dashes) from \citet[light and dark gray curves]{muzzin13b} and \citet[blue and purple curves]{tomczak14}, and their total 1$\sigma$ errors in the respective redshift bins (thin curves). \textit{Top panels:} The fractional correction to account for close-pairs on the SMF for the redshift and stellar mass bins considered ($\Phi_{\sc{corr}}/ \Phi_{\rm{old}}$). } \label{fig3-smf}
\end{figure*}

To investigate the effect of blending on the high-mass end tail of the measured SMF, we estimated the "blending correction" factor necessary to the number density of observed galaxies to account for this effect. Specifically, we compared the number of galaxies in the \textit{HST} sample before and after correcting for blending in redshift bins of $z=1.5-2.5,~2.5-3.0$ and in stellar mass bins of $\log(M_{*} /M_{\odot} )= 11.00 - 11.25, 11.25 - 11.50 $ and $> 11.50$. We applied this factor to the volume density of galaxies above the completeness limit for each survey and field in identical $M_{*}$ and $z$ bins. 

Figure~\ref{fig3-smf} shows the calculated SMFs at $1.5<z<2.5$ and $2.5<z<3.0$ before (black stars) and after correcting (red stars) for the effect of galaxy blending in the ground-based $K$~band imaging. Also overplotted are SMFs at the targeted redshifts from \citet[light and dark gray curves]{muzzin13b} and \citet[blue and purple curves]{tomczak14} in their probed stellar mass regimes. We find that at $2.5<z<3.0$, blending in ground-based K-band imaging does not seem to significantly effect the extreme massive end ($\log(M_{*}/M_{\odot})>11$) of the SMF. However, at $1.5<z<2.5$, the effect of blending is substantial for the largest stellar mass bin considered, at the level of a factor $\sim1.5$. We note that the blending-corrected results are consistent with the SMFs of \citet{muzzin13b}.

\begin{figure*}
\epsscale{0.9}
\plotone{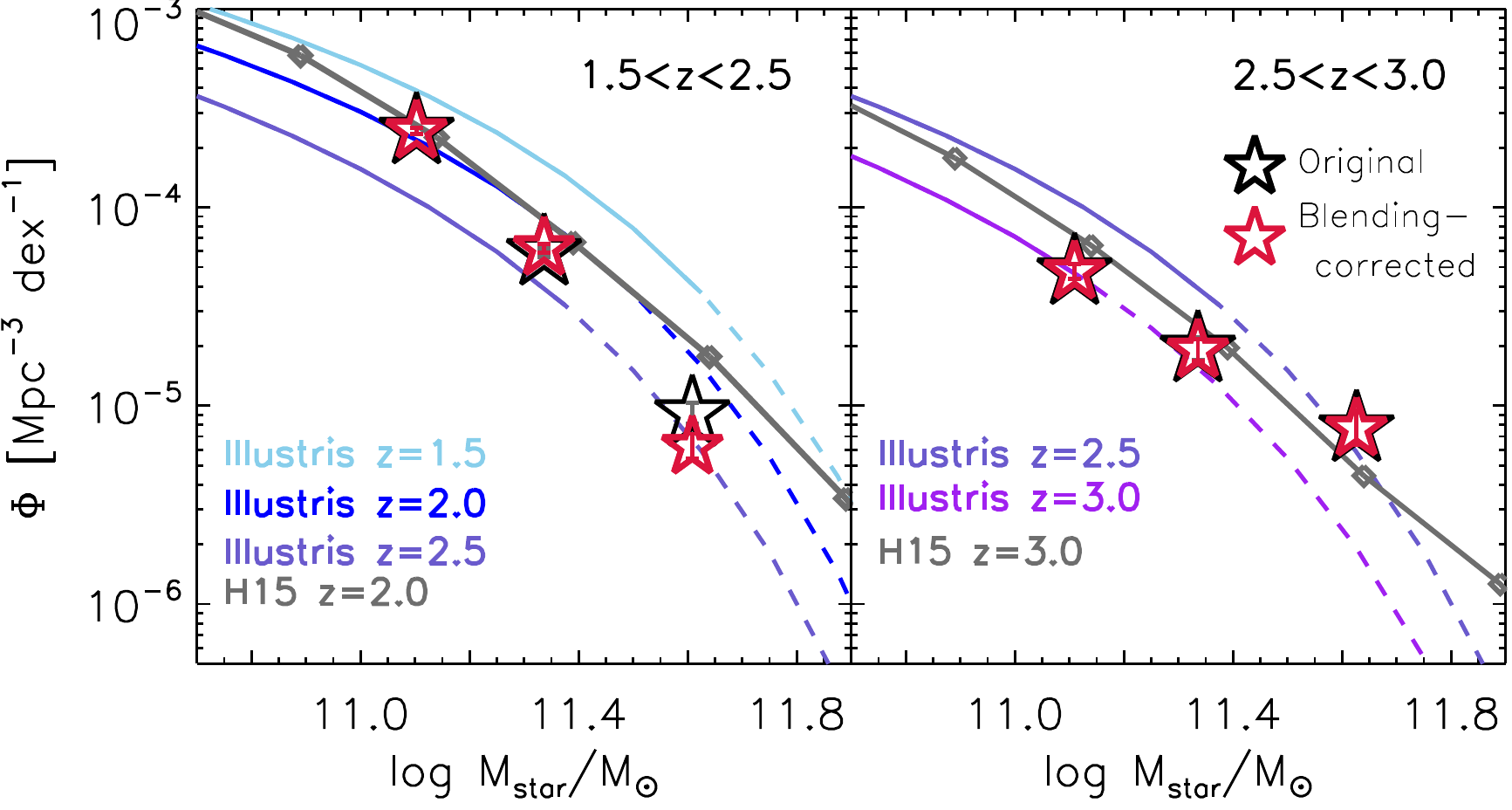}
\caption{The stellar mass function, with star symbols identical to those in Figure~\ref{fig3-smf}. Gray curves represent the $z=2$ and $z=3.0$ stellar mass functions from the Munich galaxy formation model presented in \citet{henriques15}. 
Blue solid curves indicate galaxy stellar mass functions from the Illustris cosmological hydrodynamical simulation \citep{torrey15}, calculated at the limiting redshifts for each panel ($z=1.5, 2.0, 2.5, 3.0$ mass functions shown in light blue, blue, dark blue and purple, respectively; dashed curves indicate the extrapolation of the mass function to $\Phi<3\times 10^{-5}$Mpc$^{-3}$dex$^{-1}$). } \label{fig3-smfmod}
\end{figure*}

Figure~\ref{fig3-smfmod} shows the same calculated number densities as Figure~\ref{fig3-smf}, this time compared with the predictions from theoretical studies. The gray solid curve in each redshift panel represents the galaxy stellar mass function from 
the updated Munich galaxy formation model of \citet{henriques15} (with stellar masses shifted by +0.14 dex to convert from \citealt{maraston05} to \citealt{bc03} stellar populations), calculated based on the Millennium \citep{springel05} and Millennium-II \citep{boylankolchin09} dark matter simulations updated to \textit{Planck} first-year cosmology. 
Also shown in each panel are the fits to the (differential) galaxy stellar mass functions from the cosmological hydrodynamical simulation Illustris \citep{torrey15}, calculated at the limits of each redshift bin. The Illustris SMFs at $z=1.5, 2.0, 2.5, 3.0$  
are shown in light blue, blue, dark blue and purple solid curves respectively, for the valid range indicated in that study ($\Phi>3\times 10^{-5}$Mpc$^{-3}$dex$^{-1}$; the extrapolation of the fit to lower mass function values is indicated in dashed curves). 

We find that the predicted SMFs presented in \citet{torrey15} for galaxies at $1.5<z<3.0$ are consistent with observations within the valid parameter space of Illustris. 
The volume probed by Illustris limits inferring predictions for the mass function at the extreme massive end of the galaxy population due to the rarity of these objects, which corresponds to a lower limit on the value of the stellar mass function ($\Phi=3\times 10^{-5}$ Mpc$^{-3}$ dex$^{-1}$, corresponding to $\log(M_{*}/M_{\odot}$)$\sim11.6$ and 11.1 at $z\approx1.5$ and 3.0, respectively). We highlight the large volumes necessary to make predictions for the extraordinary, ultra-massive galaxies at $z>1.5$. Specifically, next generation of hydrodynamical simulations such as \textsc{IllustrisTNG} \citep{pillepich17} is necessary to infer the behavior of the SMF at the extreme massive end of galaxy populations. 

As expected from the larger volumes of SAMs, predictions from \citet{henriques15} probes the SMF to larger stellar masses. 
In the low redshift bin ($1.5<z<2.5$), \citet{henriques15} over-predicts the abundance of galaxies. We note that this is also the stellar mass bin where galaxy blending most significantly affects the inferred number density of galaxies, increasing the tension between theoretical predictions and observations. The discrepancy between \citet{henriques15} predictions and our observations is more evident in the higher redshift bin ($2.5<z<3.0$), where the SMF is underestimated for galaxies at $\log(M_{*}/M_{\odot})>11.4$. In contrast to the trend observed at $z<2.5$, the effect of galaxy blending works to bring the observed SMF more in line with theoretical predictions, although the estimated correction for blending is negligible in this redshift bin.
However, we note that the remaining disagreement is not significant after accounting for uncertainties due to SED-modeling assumptions (a potential factor of $\sim2$, i.e., $\sim0.3$~dex in stellar mass).
This makes it clear that deriving accurate $M_{*}/L_{H}$ ratios, and hence, stellar masses for these targets is necessary through detailed spectroscopic analyses.

Finally, we stress that this early investigation of the effects of blending on the inferred number densities of very massive galaxies at high redshift serves to illustrate that this is an additional avenue to rein in on the systematic uncertainties related to observationally characterizing this population. More to the point, we caution that the inferred correction factors should not be blindly applied to different datasets.
\section{Discussion and Summary} \label{sec3-summary}

We presented the investigation of the structural properties of very massive galaxies ($\log(M_{*}/M_{\odot})>11.2$) at $1.5<z<3.0$. 
Owing to their low spatial density in the distant universe, identifying and assembling a large enough sample of very massive galaxies requires large survey volumes. We selected a sample of 37 galaxies from the combined UltraVISTA, NMBS-II and UDS catalogs to perform \textit{HST} WFC3/F160W follow-up imaging in order to accurately determine their sizes and morphologies. We modeled their 2D light profiles using GALFIT and compared their size distributions with the high-$z$ sample of \citet{vanderwel14a}. Visual investigation of the $H_{160}$ imaging revealed that 13/37 targets were unresolved in the parent $K$~band catalogs. 
We investigated the effect of galaxy blending on the SMF at $1.5<z<3.5$ by decomposing the estimated stellar masses of the close-pair systems based on their observed $H_{160}$ fluxes.
Based on this analysis the results can be summarized as follows:
\begin{itemize}
\item{At $1.5<z<2.5$, the sizes of both star-forming and quiescent galaxies with $\log(M_{*}/M_{\odot})>11.2$ are relatively consistent with those found in \citet{vanderwel14a}.}
\item{At $2.5<z<3$, sizes for quiescent galaxies at $\log(M_{*}/M_{\odot})>11.2$ appear to be systematically larger than what is expected based on the extrapolation of the relation derived from lower stellar mass galaxies, confirming results in \citet{patel17}. }
\item{We found that the effect of galaxy blending is most significant for the largest stellar mass bin ($\log(M_{*}/M_{\odot})\approx11.6$) considered at $1.5<z<2.5$, although it remains consistent with the SMF of \citet[as calculated from the maximum likelihood method]{muzzin13a}. }
\item{From the comparison with theoretical predictions, we find that the Illustris simulation agrees well with the observed number density, although their simulated volume is too small to probe the most massive galaxies. Similarly good agreement at $\log(M_{*}/M_{\odot}$)$<11.5$ is found between observations and the predictions from the SAM of \citet{henriques15}. However, the observed number density of the most massive galaxies (i.e., $\log(M_{*}/M_{\odot}$)$>11.5$) is over-predicted at $1.5<z<2.5$ and under-predicted at $2.5<z<3.0$.}

\end{itemize}

\acknowledgements
 Z.C.M. gratefully acknowledges support from the Faculty of Science at York University as a York Science Fellow. D.M. and Z.C.M acknowledge support from the program HST-GO-12990 provided by NASA through a grant from the Space Telescope Science Institute, which is operated by the Association of Universities for Research in Astronomy, Incorporated. 
The Cosmic Dawn Center is funded by the Danish National Research Foundation.

 \facility{HST(WFC3)}
 
 \bibliographystyle{aasjournal}
\bibliography{references}

\end{document}